\documentclass{mn2e}

\usepackage{amsfonts}
\usepackage{amsmath}
\usepackage{graphicx}

\title[Galaxy colors and clustering]
      {A halo model of galaxy colors and clustering in the SDSS}
\author[R. A. Skibba \& R. K. Sheth]
 {Ramin A. Skibba$^1$ \& Ravi K. Sheth$^2$
  \thanks{E-mail:  skibba@mpia.de (RAS); shethrk@physics.upenn.edu (RKS)}\\
  $^{1}$Max-Planck-Institute for Astronomy, K\"{o}nigstuhl 17,
	D-69117 Heidelberg, Germany\\
  $^{2}$Department of Physics \& Astronomy, University of Pennsylvania,
	209 S. 33rd Street, Philadelphia, PA 19130, USA}

\newcounter{appfig}

\begin{document}

\pagerange{\pageref{firstpage}--\pageref{lastpage}}

\label{firstpage}

\maketitle

\begin{abstract}
Successful halo-model descriptions of the luminosity dependence of 
clustering distinguish between the central galaxy in a halo and 
all the others (satellites).  To include colors, we provide a 
prescription for how the color-magnitude relation of centrals 
and satellites depends on halo mass.  This follows from two assumptions:  
 (i) the bimodality of the color distribution at fixed luminosity 
     is independent of halo mass, and 
 (ii) the fraction of satellite galaxies which populate the red 
      sequence increases with luminosity.
We show that these two assumptions allow one to build a model of 
how galaxy clustering depends on color without any additional free 
parameters than those required to model the luminosity dependence 
of galaxy clustering.  We then show that the resulting model is in 
good agreement with the distribution and clustering of colors in 
the SDSS, both by comparing the predicted correlation functions 
of red and blue galaxies with measurements, and by comparing the 
predicted color mark correlation function with the measured one.  
Mark correlation functions are powerful tools for identifying and 
quantifying correlations between galaxy properties and their  
environments:  our results indicate that the correlation between 
halo mass and environment is the primary driver for correlations 
between galaxy colors and the environment; additional correlations 
associated with halo `assembly bias' are relatively small.
Our approach shows explicitly how to construct mock catalogs which 
include both luminosities {\em and} colors --- thus providing 
realistic training sets for, e.g., galaxy cluster finding algorithms.  
Our prescription is the first step towards incorporating the entire 
spectral energy distribution into the halo model approach.  
\end{abstract}


\section{Introduction}
The halo model is a useful language for discussing how galaxy 
clustering depends on galaxy type:  galaxy bias (see Cooray \& Sheth 2002 
for a review).  To date, the halo model has been used to provide 
a useful framework for modeling the luminosity dependence of galaxy 
clustering.  The main goal of this paper is to extend the halo 
model description of galaxy luminosities to include colors.  
This is an important step towards the ultimate goal of providing a 
description of how the properties of a galaxy, its morphology and 
spectral energy distribution, are correlated with those of its neighbors.  
The hope is that, by relating such correlations between galaxies to 
the properties of their parent dark matter halos, the halo model 
will provide a useful guide in the study of galaxy formation.  

The halo model description of the luminosity dependence of clustering 
is usually done in three rather different ways, which have come to be 
known as the `halo occupation distribution' (HOD;  
Jing, Mo \& B\"orner 1998; Benson et al. 2000; Seljak 2000; 
Scoccimarro et al. 2001; Berlind \& Weinberg 2002; Zehavi et al. 2005) 
the `conditional luminosity function' 
(CLF; Peacock \& Smith 2000; Yang et al. 2003; Cooray 2006; 
van den Bosch et al. 2007a), 
and the `subhalo abundance matching' (SHAM; Klypin et al. 1999; 
Kravtsov et al. 2004; Vale \& Ostriker 2006; Conroy, Wechsler \& 
Kravtsov 2006) methods.  
The HOD approach uses the abundance and spatial distribution of a 
given galaxy population (typically, just the two-point clustering 
statistics) to determine how the number of galaxies depends on the 
mass of the parent halo.  This is done by studying a sequence of 
volume limited galaxy catalogs, each containing galaxies more luminous 
than some threshold luminosity.  
The CLF method attempts, instead, to match the observed luminosity 
function by specifying how the luminosity distribution in halos 
changes as a function of halo mass.  
One can infer the CLF from the HOD approach, and vice-versa, so the 
question arises as to which is the more efficient description.  For 
a given catalog, the HOD method requires the fitting of just two 
free parameters, so it is relatively straightforward.  
The CLF method requires many more parameters to be fit simultaneously, 
but uses fewer volume limited catalogs.  
SHAMs first identify the subhalos within virialized halos in 
simulations, and then use subhalo properties to match the subhalo 
abundances to the observed distribution of luminosities.  Once this 
has been done, CLFs or HODs can be measured in the simulations.  

In SPH and semi-analytic galaxy formation models, central and 
satellite galaxies are rather different populations 
(\textit{e.g.}, Kauffmann et al. 1999; Sheth \& Diaferio 2001; 
Guzik \& Seljak 2002; Benson et al. 2003; Sheth 2005; 
Zheng et al. 2005).  And so too, in the HOD and CLF approaches 
to the halo model, the central galaxy in a halo is assumed to be very 
different all the others, which are called satellites.  For example, 
the CLF approach must provide a description of how the central and 
satellite luminosity functions vary as a function of halo mass.  
The HOD-based analyses predict that the satellite galaxy luminosity 
function should be approximately independent of halo mass, and hence 
of group and/or cluster properties (Skibba et al. 2006).
Skibba et al. (2007) present evidence from the SDSS in support of 
this prediction.  More recent analysis of a rather different group 
catalog has confirmed this finding (Hansen et al. 2008).
Skibba et al. argued that this independence can reduce the required 
number of free parameters in CLF-based analyses.  

One of the goals of the present work is to show that the HOD-based 
approach also provides a rather simple way to understand how galaxy 
clustering depends on color.  In essence, it provides a simple 
algorithm for specifying how the joint CLF (\textit{i.e.}, 
the luminosity distribution in two different bands) varies with 
halo mass.  In principle, this can be done by splitting the sample 
up into small bins of luminosity {\em and} color, and studying how 
the clustering signal in each bin changes.  Zehavi et al. (2005) 
describe a first attempt at this -- for each bin in luminosity, they 
use two bins in color: `red' or `blue'.  (Croton et al. 2007 also 
study the difference in clustering strengths of red and blue 
galaxies.  They use related statistics, but do not attempt a 
halo-model description of their measurements.)  As sample sizes increase, 
it will become possible to split the sample into many more color 
bins.  However, even for this simplest case, Zehavi et al. were led 
to a rather more complex parametrization of the HOD than was necessary 
for the luminosities -- they caution that, as a result, there are more 
degeneracies amongst their parameter choices, and so the constraints 
on the HODs they obtain are considerably weaker than for luminosities 
alone.  While such a brute force approach to determining the HOD is 
certainly possible, we argue below that there may be some merit to 
recasting the problem as one in which the physics and statistics are 
more closely related.  

In essence, our approach exploits the fact that, to a good 
approximation, galaxies appear to be bimodal in their properties 
(\textit{e.g.}, Blanton et al. 2003).  
In the present context, we are interested in the fact that the 
distribution of colors at fixed luminosity is bimodal (\textit{e.g.}, 
Baldry et al. 2004; Willmer et al. 2006). 
Our approach is to couple this 
bimodality with the centre-satellite split in the halo model.  

This paper is organized as follows.  Section~\ref{model} describes 
our approach:  it shows the correlation between color and luminosity 
in the SDSS sample, and then describes a model for the luminosities 
and colors of centrals and satellites which is designed to reproduce 
this bimodality.
Section~\ref{twotests} describes how to use our model to generate 
mock catalogs which have the correct luminosity dependence of 
clustering and the observed color-magnitude relation, as well as 
how to incorporate our approach into a halo model description of 
the color-mark two-point correlation function.
Section~\ref{compare} provides a comparison of our model predictions 
with measurements from the SDSS.
These include the clustering signal from `red' and `blue' galaxies 
(defined as being redder or bluer than a critical luminosity dependent 
color) and the clustering signal when galaxies are weighted by 
color -- the color-mark correlation function.  
A final section summarizes our findings.  

Throughout, the restframe magnitudes we quote are associated 
with SDSS filters shifted to $z=0.1$; the absolute magnitude of 
the Sun in this $r$-band filter is 4.76 (Blanton et al. 2003).  
Where necessary, we assume a flat background cosmological 
model in which $\Omega_0=0.3$, the cosmological constant 
is $\Lambda_0=1-\Omega_0$, and $\sigma_8=0.9$.  We write the
Hubble constant as $H_0=100h$~km~s$^{-1}$~Mpc$^{-1}$.
In addition, we always use `log' for the 10-based logarithm
and `ln' for the natural logarithm.

\section{Color-magnitude bimodality and 
         the centre-satellite split}\label{model}

\subsection{Bimodality in the SDSS}\label{sdss}

\begin{figure}
 \centering
 \includegraphics[width=\hsize]{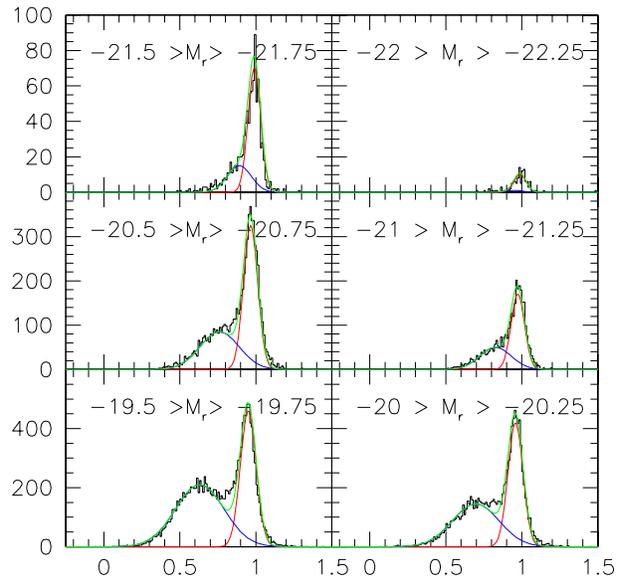} 
 \caption{Bimodal distribution of $g-r$ color in the SDSS.  Smooth 
          curves show that, at fixed luminosity, the distribution is 
          well modeled by the sum of two Gaussian components.}
 \label{bimodal}
\end{figure}

Baldry et al. (2004) report that the distribution of rest-frame 
$u-r$ color at fixed $r$-magnitude can be well-modeled as the 
sum of two Gaussian components.  The same is true of the 
distribution of rest-frame $g-r$ color (\textit{e.g.}, Blanton et al. 2005); 
we call these the red and blue components of the distribution $p(c|L)$.  
The mean and rms values of these components depend on luminosity. 
This dependence is quite well described by simple power laws:  
\begin{eqnarray}
 \label{reds}
 \Bigl\langle g-r|M_r\Bigr\rangle_{\rm red} &=& 0.932 - 0.032\,(M_r+20), 
 \nonumber\\
 {\rm rms}\Bigl(g-r|M_r\Bigr)_{\rm red} &=& 0.07 + 0.01\,(M_r+20);\\
 \Bigl\langle g-r|M_r\Bigr\rangle_{\rm blue} &=& 0.62 - 0.11\,(M_r+20), 
 \nonumber\\
 {\rm rms}\Bigl(g-r|M_r\Bigr)_{\rm blue} &=& 0.12 + 0.02\,(M_r+20).
 \label{blues}
\end{eqnarray}
The fraction of objects in the blue component decreases with 
increasing luminosity:
\begin{equation}
 f_{\rm blue}(M_r) \approx 0.46 + 0.07\,(M_r+20), 
 \label{fblue}
\end{equation}
and drops toward zero at the bright end.

\begin{figure}
 \includegraphics[width=\hsize]{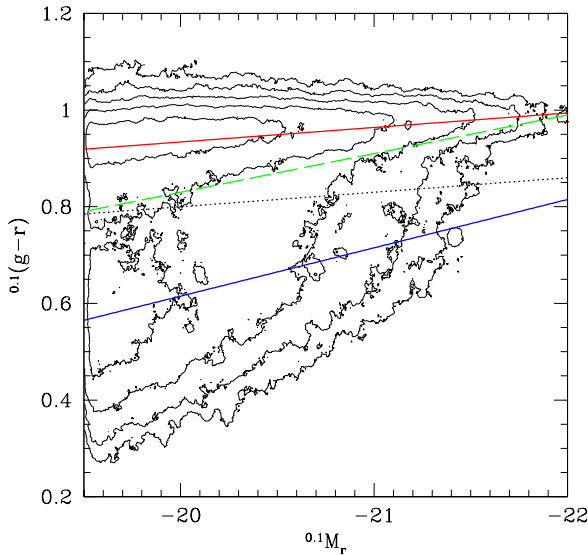} 
 \caption{Color-magnitude diagram in the $M_r<-19.5$ volume-limited 
          SDSS catalog.
          Solid lines show the mean values of the red and blue 
          sequences (equations~\ref{reds} and~\ref{blues}); 
          dashed line shows the satellite sequence (equation~\ref{Csatseq}), 
          and dotted line shows equation~(\ref{colorcut}) which some 
          authors use to divide the population into red and blue.}
 \label{Mr195CMD}
\end{figure}

Figure~\ref{bimodal} shows this bimodality, and the two 
Gaussian component fits which are based on these expressions.  
Our model of the bimodality, which 
motivates an algorithm for constructing mock catalogs, and which 
our halo model calculation requires, uses the red and blue sequences 
given by equations~(\ref{reds}) and~(\ref{blues}).  
These sequences are also shown in a color-magnitude diagram,
Figure~\ref{Mr195CMD}, along with the color-magnitude contours of one 
of the volume-limited SDSS catalogs used in Section~\ref{compare}.

However, it is common to make a cruder approximation to this 
bimodality, by simply labeling galaxies as `red' if they are 
redder than 
\begin{equation}
  {^{0.1}(g-r)}_\mathrm{cut} \,=\, 0.8 \,-\, 0.03\,({^{0.1}M_r}+20),
  \label{colorcut}
\end{equation}
and calling them `blue' otherwise (\textit{e.g.}, Zehavi et al. 2005; 
Blanton \& Berlind 2007).  (The recent analysis of satellite galaxy 
colors by van den Bosch et al. 2007b used a stellar mass-based split, 
which translates into a similar color cut as the one above, although 
their cut is slightly steeper with respect to $r$-band luminosity.)  
In what follows, we will only use this sharp threshold when comparing 
our results to previous work.  

The SDSS colors (and magnitudes) have measurement errors which 
contribute to the rms of the red and blue sequences, especially 
at faint magnitudes.  However, the uncertainties in the $g-r$ galaxy 
colors in the SDSS are typically less than 0.02~mags, so they are 
unlikely to significantly affect the constraints on the model.  
Since the measurement errors almost certainly do not correlate with 
environment, they are not expected to bias the measured color mark 
correlation functions shown in Section~\ref{compare}; they will,  
however, increase the error bars on the clustering signal.  
We note, however, that there is an important systematic problem with 
the colors for which we do not correct:  namely, a dusty spiral will 
appear redder if viewed edge-on rather than face-on.  In fact, a 
significant fraction of the objects called `red' are not the 
early-types which one typically associates with the `red sequence' 
(Bernardi et al. 2003).  Mitchell et al. (2005) estimate that this 
fraction is of order 40\% (also see Maller et al. 2008).  Since this 
systematic also affects the luminosities, for which no halo-model 
analysis to date has yet made a correction, we have not done so here 
either.  

\subsection{Luminosities and colors of centrals and satellites}\label{cssplit}

To illustrate our approach we will begin with an extreme assumption.  
Suppose that: (i) the bimodal color distribution is independent of halo 
mass (by which we mean that the distribution of color at fixed 
luminosity is independent of halo mass; the distribution of 
luminosities, of course, does depend on halo mass), and that 
(ii) satellites are drawn from the red part of the bimodal color 
distribution -- {\em no} satellites come from the blue sequence.  
Later in this paper, we will find it necessary to relax the second 
assumption, but the data does not yet require us to give up the first.  
We think assumption (ii) is a useful extreme which helps bring into 
focus the key points of the approach.  

Given the constraints from the color distribution as a function 
of luminosity (Section~\ref{sdss}) and from luminosity-dependent 
clustering (Appendix~\ref{zzHOD}), these two assumptions allow one to 
model the halo mass dependence of the colors of both centrals and 
satellites, and in general to build a model of how galaxy clustering 
depends on color, without any additional free parameters. 
For example, these assumptions imply that the mean satellite color is 
\begin{eqnarray}
 \langle c|m\rangle &\equiv& \int dc\, p(c|m)\, c 
                    = \int dL\, p(L|m) \int dc\, p(c|L,m)\, c \nonumber\\
                    &=& \int dL\, p(L|m) \langle c|L,m\rangle .
\end{eqnarray}
Whereas the first equality is the definition, the final expression 
shows how one might estimate the left-hand side from a knowledge 
of the luminosity distribution in halos of mass $m$ and the mean 
color at given luminosity in such halos.  

If the distribution of satellite colors at fixed satellite luminosity 
is independent of halo mass (this is not unreasonable, given that 
the distribution of luminosities themselves is approximately 
independent of halo mass; see Skibba et al. 2006, 2007; 
Hansen et al. 2008), then this becomes 
\begin{equation}
 \langle c|m\rangle_{\rm sat} = \int dL\, p_{\rm sat}(L|m)\, 
                                \langle c|L\rangle_{\rm sat} .
 \label{cmsatint}
\end{equation}
Thus, given $m$, we integrate over the distribution of satellite 
luminosities, weighting by $\langle c|L\rangle_\mathrm{sat}$.  

Our simplest model (assumption ii) uses equation~(\ref{reds}), 
the color magnitude relation along the red sequence, for 
$\langle c|L\rangle_\mathrm{sat}$.  We will show later that setting
\begin{equation}
 \langle g-r|M_r\rangle_\mathrm{sat} = 0.83 - 0.08\,(M_r+20)
 \label{Csatseq}
\end{equation}
instead, which is bluer at faint luminosities (see Figure~\ref{Mr195CMD}),
provides substantially better agreement with the observations.  
This is best thought of as a model in which satellites are drawn 
from the red sequence with probability 
\begin{equation}
 p({\rm red\ sat}|L) = 
  \frac{\langle c|L\rangle_\mathrm{sat} - \langle c|L\rangle_\mathrm{blue}}
       {\langle c|L\rangle_\mathrm{red} - \langle c|L\rangle_\mathrm{blue}},
 \label{predsatL}
\end{equation}
and from the blue sequence with probability 
\begin{equation}
 p({\rm blue\ sat}|L) = 1 - p({\rm red\ sat}|L).
 \label{pbluesatL}
\end{equation}
These expressions imply that, for SDSS $g-r$ colors, 
$p({\rm blue\ sat}|L) \approx 0.4$ at $M_r = -18$, and it drops 
to zero at $M_r\approx -22$.  
Since the fraction of galaxies that are satellites has a similar 
dependence on luminosity (we provide explicit HOD-derived expressions 
for this later), this model says that although almost sixty percent 
of the galaxies with $M_r=-18$ are from the blue sequence 
(c.f. equation~\ref{fblue}), slightly less than twenty percent of 
the galaxies with $M_r = -18$ are blue satellites:  only a third 
of the faint blue galaxies are satellites, the others are centrals.  
Allowing for blue-sequence satellites modifies the discussion below 
trivially.  

It is worth reiterating that, in this model, satellite colors only 
depend on halo mass because satellite luminosities do.  Since 
$p_{\rm sat}(L|m)$ depends only weakly on $m$ (Skibba et al. 2007), 
we expect $\langle c|m\rangle_{\rm sat}$ to also depend only weakly 
on $m$.  

In practice, we do not evaluate the integral in equation~(\ref{cmsatint}) 
as written.  Rather, we use a variation of the trick we used in 
Skibba et al. (2006).  Namely, for some function $C(L)$ of $L$, 
\begin{equation}
 \int_{L_{\rm min}}^\infty \!\!\! dL\,C(L) \int_L^\infty dL'\,p(L'|m)
 = \int_{L_{\rm min}}^\infty \!\!\! 
    dL'\,p(L'|m)\,\int_{L_{\rm min}}^{L'} dL\,C(L).
\end{equation}
Skibba et al. studied the case where $C(L)=1$, so the inner integral 
gave $L'-L_{\rm min}$.  Here, we wish to set $C(L)$ to be that 
function of $L$ which, when integrated over $L$, yields 
$\langle c|L'\rangle_{\rm red} - \langle c|L_{\rm min}\rangle_{\rm red}$.  
Thus,  
\begin{equation}
 \langle c|m\rangle_{\rm sat} = \langle c|L_{\rm min}\rangle_{\rm red} 
 + \int_{L_{\rm min}}^\infty \!\!\! dL\,C(L)\, P_{\rm sat}(>L|m),
\end{equation}
where we have defined 
\begin{equation}
 P_{\rm sat}(>L|m) \equiv \int_L^\infty dL\,p_{\rm sat}(L|m) 
  = {N_{\rm sat}(>L|m)\over N_{\rm sat}(>L_{\rm min}|m)}.
\end{equation}
If color and luminosity are in magnitudes (\textit{i.e.}, we work in 
logarithmic rather than linear variables) then the integral is simpler:
\begin{eqnarray}
 \langle g-r|m\rangle_{\rm sat} &=& \langle g-r|M_{\rm min}\rangle_{\rm sat} +
  \nonumber\\
 && C_{{\rm sat},\,{\rm slope}} \,
       \int_{M_{r, \rm min}}^{-\infty} \!\!\! dM_r\,P_{\rm sat}(<M_r|m),
\end{eqnarray}
where $P_{\rm sat}(<M_r|m) = P_{\rm sat}(>L|m)$, and 
$C_{\rm sat,\ slope}$ is the slope of the relation showing how the 
mean satellite color changes with magnitude.  That is, 
$C_{\rm sat,\ slope} = -0.032$ or $-0.08$ if satellites are 
drawn from the red sequence (c.f. equation~\ref{reds}) or from 
equation~(\ref{Csatseq}).

Obtaining an expression for the typical color associated with the 
central galaxies of $m$ halos is more complicated.  
Although the bimodal distribution of color at fixed luminosity 
can be thought of as arising from a mix of objects which lie 
along a blue or a red sequence, in what follows, it will be 
more useful to think in terms of the central-satellite split.  
In this case, 
\begin{equation}
 \langle c|L\rangle = {N_{\rm cen}(L) \langle c|L\rangle_{\rm cen} + 
                     N_{\rm sat}(L) \langle c|L\rangle_{\rm sat}\over 
                      N_{\rm cen}(L) + N_{\rm sat}(L)}
\end{equation}
making 
\begin{equation}
  \langle c|L\rangle_{\rm cen} = \langle c|L\rangle + 
 {N_{\rm sat}(L)\over N_{\rm cen}(L)}\,
 \Bigl[\langle c|L\rangle - \langle c|L\rangle_{\rm sat}\Bigr].
 \label{Ccen}
\end{equation}
If, as we assumed for the satellites, the distribution of central 
galaxy colors at fixed luminosity is independent of halo mass 
(the results of Berlind et al. 2005 support this assumption),
then the mean color as a function of halo mass is simply
$\langle c|m\rangle_{\rm cen}\,=\,\langle c|L(m)\rangle_{\rm cen}$
if there is no scatter between central galaxy luminosity and halo 
mass (\textit{e.g.}, Zehavi et al. 2005).  If there is scatter 
(\textit{e.g.}, Zheng et al. 2007), then 
\begin{equation}
  \langle c|m\rangle_{\rm cen} 
  = \int dL\, P_{\rm cen}(L|m)\,\langle c|L\rangle_{\rm cen}.
 \label{CcenM}
\end{equation}

Now, by hypothesis, $\langle c|L\rangle_{\rm sat}$ is given by 
equation~(\ref{reds}) (or equation~\ref{Csatseq}), whereas 
$\langle c|L\rangle$ is simply the mean color of all galaxies 
as a function of luminosity.  
Thus, both these quantities are observables, or are constrained 
by observables, for the satellites (Skibba 2008); 
the only unknown is $N_{\rm sat}(L)/N_{\rm cen}(L)$.  
Since both numbers are counted in the same volume, this is 
the same as the ratio of the number densities:  
$n_{\rm sat}(L)/n_{\rm cen}(L)$.  We discuss how this ratio is 
determined by the luminosity-based HOD in Appendix~\ref{zzHOD}.  

It is worth noting that the quantity in square brackets in 
equation~(\ref{Ccen}) is negative. 
This means that, in general, the colors of central galaxies are 
{\em bluer} than the average for their luminosities.  
Although this seems counter to intuition---one is used to thinking 
of central galaxies as being red---it is, in fact, sensible.  
Essentially, the paradox is resolved when one realizes that the 
satellites actually inhabit more massive halos than do centrals 
of the same luminosity.  
It may help to note that this effect is most pronounced at low $L$, 
where the mean color is significantly bluer than the red sequence, 
and the number of satellites can be large.  Low luminosity 
galaxies that are centrals are hosted by low mass halos, 
whereas satellite galaxies of similar luminosity are more likely to 
reside in groups or clusters, so their parent halos are more massive.  
Thus, our model has placed blue central galaxies in low mass halos 
and red satellite galaxies in massive halos.  
At higher luminosities, $\langle c|L\rangle$ approaches that of the 
red sequence.  In this limit, the term in square brackets becomes 
small, as does the number of satellites, so the colors of central 
galaxies tend to $\langle c|L\rangle$:  that is, our model places 
luminous central galaxies on the red sequence.  

\section{Two ways to test the model of bimodality}\label{twotests}
We now describe two ways to test our model of the bimodality.  
The first is numerical --- we provide an algorithm for constructing 
mock catalogs which are consistent with our model.  The model can 
be tested by performing the same analysis on the mocks that was 
performed on real data.  This is particularly useful for analyses 
which are somewhat involved or contrived, so that an analytic 
description is difficult.  
The second is analytic --- we show how our model can be implemented 
to provide a halo model description of mark correlations when the 
mark is color.  
Skibba (2008) describes the result of a third test:  a direct 
measurement of central and satellite colors in group catalogs.  

\subsection{An algorithm for constructing mock catalogs with 
            luminosities and colors}\label{mock}

The analysis above shows that one can generate a mock galaxy catalog 
in two steps:  first generate luminosities, and then use them to 
generate colors.  Note that the method used for generating 
luminosities is {\em not} important:  the luminosities could have 
come from an HOD analysis, a CLF analysis, or they may be based on 
a SHAM.  

Our algorithm for generating luminosities comes from 
Skibba et al. (2006).  Briefly, we specify a minimum luminosity 
$L_{\rm min}$ which is smaller than the minimum luminosity we wish 
to study.  
We then select the subset of halos in the simulation which have 
$m>m_{\rm min}(L_{\rm min})$.  Each halo is assigned a central 
galaxy with luminosity given by inverting the relation between 
halo mass and luminosity (equation~\ref{Lcen}).  
We specify the number of satellites the halo contains by 
choosing an integer from a Poisson distribution with mean 
$N_{\rm sat}(>L_{\rm min}|m)$.  The luminosity of each satellite 
galaxy is specified by generating a random number $u_0$ distributed 
uniformly between 0 and 1, and finding that $L$ for which 
$N_{\rm sat}(>L|m)/N_{\rm sat}(>L_{\rm min}|m) = u_0$.  This ensures 
that the satellites have the correct luminosity distribution.  

We could assign colors to each of the satellites by drawing a Gaussian 
random number with mean and rms given by inserting the satellite 
luminosity in equation~(\ref{reds}) for the red sequence.
However, as we show in Section~\ref{compare}, this results in a 
correlation between color and environment that is too strong compared 
to the data.  Instead, we want the satellites to have colors which are 
bluer than the red sequence at faint luminosities, as specified by 
equation~(\ref{Csatseq}).  
To implement this in our mock catalog, we draw a uniformly 
distributed random number $0\le u_1 < 1$.  The satellite is drawn 
from the red sequence (a Gaussian with mean and rms given by 
equation~\ref{reds}) if $u_1\le p({\rm red\ sat}|L)$, 
where $p({\rm red\ sat}|L)$ is given by equation~(\ref{predsatL})
and from the blue sequence (Gaussian with mean and rms from 
equation~\ref{blues}) otherwise.  
Note that only the luminosity matters for determining the color; 
the halo mass plays no additional role.

The colors for central galaxies can also be drawn from either the 
red or blue sequence.
To determine which, we draw another uniformly distributed random 
number $u_2$.  
If $u_2> f_{\rm blue}(L)/f_{\rm cen}(L)$, where $L$ is the central 
object's luminosity, then the object is assigned to the red sequence, 
so we draw a Gaussian with luminosity-dependent mean and rms 
given by  equation~(\ref{reds}).  Else, it is blue, and we 
use equation~(\ref{blues}) instead.  Equations~(\ref{fblue}) 
and~(\ref{fcen}) show that this assigns all central galaxies 
fainter than $M_r\approx -18.5$ to the blue sequence.  

Finally, we place the central galaxy at the center of its halo, 
and distribute the satellites around it so that they follow an 
NFW profile (see Scoccimarro \& Sheth 2002 for how this can be 
done efficiently).  
The resulting mock galaxy catalog has been constructed to have the 
correct luminosity function as well as the correct luminosity 
dependence of the galaxy two-point correlation function.  
In addition, colors in this catalog are assigned in accordance with 
the model described previously:  satellite and central galaxy colors are 
assigned such that the galaxy population as a whole has the correct
color-luminosity distribution.


\begin{figure}
 \centering
 \includegraphics[width=\hsize]{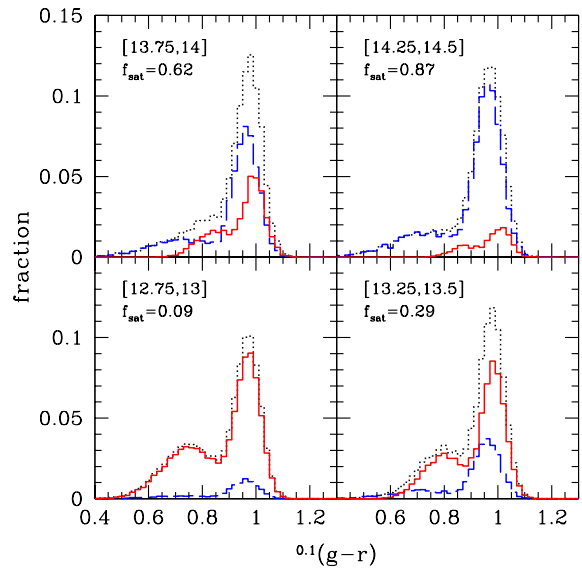}
 \caption{Bimodal distribution of $g-r$ color for central galaxies (red histogram),
          satellite galaxies (blue dashed histogram),
          and all galaxies (centrals+satellites; black dotted histogram)
          in a mock catalog with $M_r<-20.5$.
          The distributions are shown for four intervals in log halo mass, indicated
          in square brackets in each panel.}
 \label{mockgmrbimodal}
\end{figure}
Our model makes a prediction for how the bimodality in color differs for central and satellite galaxies.
In Figure~\ref{mockgmrbimodal}, we show the color distribution as a function of halo mass of
central and satellite galaxies in a mock catalog with $M_r<-20.5$.  We normalize
the central and satellite galaxy distributions by the total number of
galaxies in each bin; consequently, the lower mass halos are dominated by
central galaxies, while satellites contribute most of the galaxies in massive
halos.  First, note that the satellite distribution is almost the same in
each panel: this is a consequence of our assumption that the distribution of
satellite colors at fixed luminosity is independent of halo mass (\textit{i.e.},
$p_\mathrm{sat}(c|L,m)=p_\mathrm{sat}(c|L))$, and the fact that satellite luminosities are
approximately independent of mass as well.  On the other hand, the centrals
have a more bimodal distribution in low-mass halos, while in massive halos
most of them are on the red sequence.  Second, it is interesting that the
blue and red modes of the central galaxy bimodal color distribution are
closer together than those of the satellite color distribution, such that the
blue bump of the centrals tends to peak at the minimum in the satellite
distribution.

\subsection{Implicit assumptions, bells and whistles}
This halo-model based prescription for making mock catalogs 
uses three simplifying assumptions which are worth discussing 
explicitly.  
First, although we assume halos are spherical and smooth, the 
density run of satellites around halo centers is almost certainly 
neither.  Generating triaxial distributions is straightforward once 
prescriptions for how the triaxiality depends on halo mass and how 
it correlates with environment are available.  
Once these are known, they can be incorporated into the analytic 
halo-model description (Smith, Watts \& Sheth 2006).  Similarly, 
parametrizations of halo substructure can also be incorporated into
the description (Sheth \& Jain 2003).  Of course, both these types 
of correlations can be included in the mock catalog directly from 
a simulation if one simply selects the appropriate number of particles 
from the halo itself, rather than generating the profile shape 
synthetically.  
This is costly because now one needs the full particle distribution, 
rather than just the halo catalog, to generate the mock -- but note 
that it is not a problem of principle.  

Second, note that the number of galaxies in a halo, the spatial 
distribution of galaxies within a halo, and the assignment of 
luminosities all depend only on halo mass.  None of these depend on 
the surrounding large-scale structure.  Therefore, the mock catalog 
includes only those environmental effects which arise from the 
environmental dependence of halo abundances.  This point was made 
by Skibba et al. (2006); it is also true of our prescription for 
including colors.  

Third, halos of the same mass will have had a variety of formation 
histories.  Some will have assembled their mass and their galaxy 
populations more recently than others.  
Recent assembly means less time for dynamical friction, and, possibly, 
a younger stellar population.  So, at fixed halo mass, one might 
expect to find a correlation between the age of a halo and the 
galaxy population within it.  In particular, the number of galaxies 
in a halo, their luminosities and their colors may all be correlated 
with the formation history.  Our halo model description (and associated 
mock catalog) ignores all such correlations.  
To see this clearly, note that we assign luminosities and colors to 
the galaxies in a halo without regard for the number of galaxies in 
it.  
Had we used a SHAM to assign luminosities, then some of correlation 
between formation history and the galaxy population will have been 
included.  
If one is already carrying along the particle distribution from the 
simulation to construct the mock, then the next level of complication 
is to also include additional information about the merger history in 
the simulation, for use when making the mock.  

We also assign colors to satellite galaxies without explicit 
consideration of the color of the central galaxy, and we make no 
effort to incorporate color gradients within a halo into our model.  
This is mainly because the two-point statistics we study in this 
paper, weighted or unweighted, are known to be not very sensitive 
to gradients (see Sheth et al. 2001, Scranton 2002, Sheth 2005 and 
Skibba 2008 for more discussion and simple prescriptions for 
incorporating color gradients.)  

These are all interesting problems for the future (and they are 
almost certainly not independent problems!), but the measurements 
described in the next section do not require these refinements.

\subsection{A halo model description of color mark correlations}\label{halomark}

Mark correlations are an efficient way to quantify the correlation 
between the properties of galaxies and their environment 
(Sheth, Connolly \& Skibba 2005).  
The two-point mark correlation function is simply 
\begin{equation}
  M(r) \,\equiv\, \frac{1+W(r)}{1+\xi(r)},
 \label{markedXi}
\end{equation}
where $\xi(r)$ is the traditional two-point correlation function 
and $W(r)$ is the same sum over galaxy pairs separated by $r$, but 
now each member of the pair is weighted by the ratio of its mark to 
the mean mark of all the galaxies in the catalog 
(\textit{e.g.}, Stoyan \& Stoyan 1994; Beisbart \& Kerscher 2000).  
In effect, the denominator divides-out the contribution to the 
weighted correlation function which comes from the spatial 
contribution of the points, leaving only the contribution from the 
fluctuations of the marks.

In models where a galaxy's properties correlate with environment 
only because they correlate with host halo mass, but halo abundances 
correlate with environment, it is relatively straightforward to 
write down a halo model of mark correlations (Sheth 2005).  Since 
our model of central and satellite colors is precisely of this 
form, we can build a halo model of color-mark correlations.  
Appendix B provides a detailed description of how this is done.  
In principle, comparison of this prediction with measurements 
in the SDSS dataset allow a test of our approach.  

\begin{figure}
 \centering
 \includegraphics[width=\hsize]{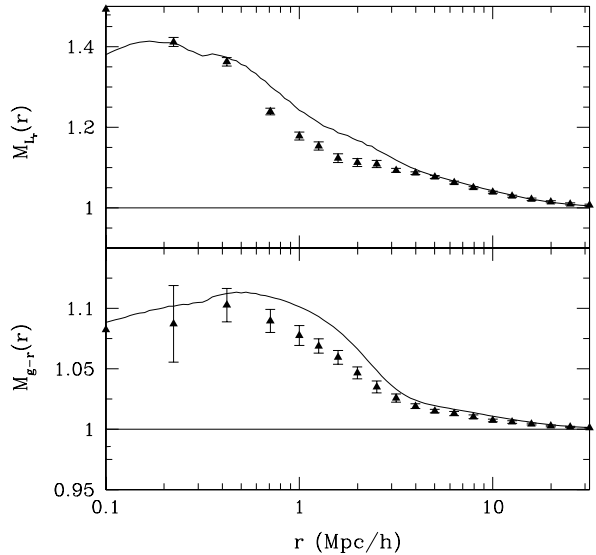} 
 \caption{Luminosity (top) and $g-r$ color (bottom) mark 
  correlation functions in a real-space mock catalog in which 
  $M_r<-20.5$.  Solid curves show the halo model predictions. 
 }
 \label{gmrMock}
\end{figure}

Before performing this test with data, Figure~\ref{gmrMock} shows 
a comparison with measurements in the mock catalog described in 
the previous section.  The halo population is from the VLS simulation 
(Yoshida et al. 2001), and the mock galaxies have $M_r<-20.5$.  
Luminosities and colors were assigned as described above:  
the top panel shows $M(r)$ as a function of real-space separation 
when luminosity is the mark; the bottom panel has $g-r$ as the 
mark.  Solid curves show the halo model prediction, computed by 
inserting the mass dependence of the mean marks for centrals and 
satellites into the mark correlation formalism of Appendix~\ref{colorMCF}.
The luminosity and color mark correlations are significantly above 
unity, which clearly shows that in denser environments we expect the
luminosities of galaxies to be brighter (top panel) and the colors 
to be redder (bottom panel).
The mark correlations also clearly show the transition from the 1-halo 
term to the 2-halo term at $r\sim\,\mathrm{Mpc}/h$, which is the virial
radius of the most massive halos at $z\sim0$.  The transition is more
pronounced than in the traditional unmarked correlation function $\xi(r)$.

There is reasonably good agreement between the halo model calculation 
and the mocks for both the luminosity and color mark correlation 
functions; the unmarked correlation functions $\xi(r)$ agree extremely 
well, so they are not shown.
Both panels in the figure show a similar but small discrepancy at 
similar scales, approximately where the 1 halo-2-halo term transition 
occurs.
Although statistically significant, this discrepancy is small 
compared to the significance with which the signal itself differs from 
unity:  the halo model calculations are qualitatively, if not 
quantitatively, correct across a wide range in scales.
The agreement between the model and the mocks is encouraging; it 
suggests that much of the environmental dependence of galaxy color 
arises from the environmental dependence of host halo mass.

\section{Comparison with SDSS}\label{compare}
In this section we compare color mark projected correlation functions
predicted by the halo model to measurements in the SDSS (York et al. 2000).  
We use two volume-limited large-scale structure samples built from the 
NYU Value-Added Galaxy Catalog (Blanton et al. 2005$b$) from SDSS DR4plus, 
which is a subset of SDSS Data Release 5 (Adelman-McCarthy et al. 2007).
We $k$-correct the magnitudes to $z=0.1$ using the 
\texttt{kcorrect v4\_1} code of Blanton \& Roweis (2007); 
the magnitudes are also corrected for passive evolution.
Our fainter catalog has limits $-23.5< {^{0.1}M_r}<-19.5$, $0.017<z<0.082$; 
it consists of 78356 galaxies with mean density 
 $\bar n_{\mathrm {gal}}=0.01061\,(h^{-1}\mathrm{Mpc})^3$.  
Our brighter catalog has
 $-23.5< {^{0.1}M_r}<-20.5$ and $0.019<z<0.125$, 
and contains 73468 galaxies with mean density 
 $\bar n_{\mathrm {gal}}=0.00280\,(h^{-1}\mathrm{Mpc})^3$.  
These luminosity thresholds approximately correspond to $M_r<M^\ast+1$ 
and $M_r<M^\ast$, where $M^\ast$ is the break in the Schechter 
function fit to the $r$-band luminosity function (Blanton et al. 2003).

\begin{figure}
 \centering
 \includegraphics[width=\hsize]{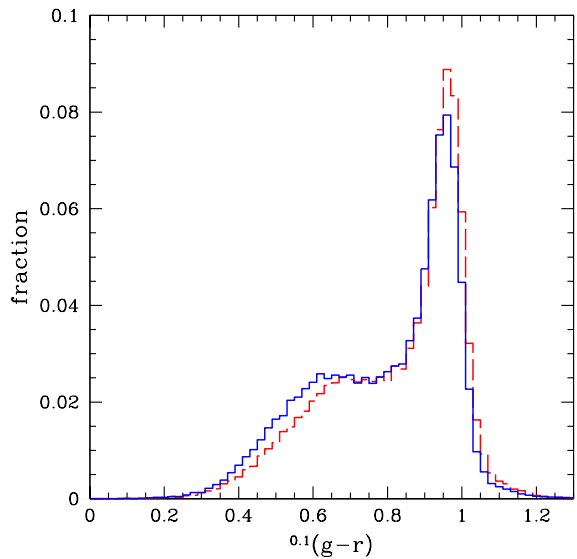}
 \caption{Distribution of Petrosian (blue histogram) and model (red dashed histogram) 
          $g-r$ colors in the $M_r<-19.5$ volume-limited catalog.}
 \label{markdist}
\end{figure}

For the measured correlation functions and jack-knife errors, which 
require random catalogs and jack-knife sub-catalogs, we use the 
hierarchical pixel scheme SDSSPix
\footnote{\texttt{http://lahmu.phyast.pitt.edu/\~scranton/SDSSPix}},
which characterizes the survey geometry, including edges and holes 
from missing fields and areas near bright stars.
This same scheme has been used for other clustering analyses
(Scranton et al. 2005, Hansen et al. 2007) and for lensing analyses 
(Sheldon et al. 2007).

Figure~\ref{markdist} shows the distribution of $g-r$ colors in our 
fainter ($M_r<-19.5$) catalog.  The distributions of Petrosian and 
model colors are similar, although the model colors are slightly 
redder.  The mean Petrosian color is 0.796, whereas the mean model 
color is 0.825.  
This is not unexpected --- galaxies have color gradients, 
and model colors measure the color on smaller scales.  
These mean values are 0.850 and 0.885 in the brighter catalog ($M_r<-20.5$).  
The blue fractions of the Petrosian colors of the fainter and brighter catalogs 
are, respectively, 44\% and 37\% using the fixed color-magnitude cut 
(equation~\ref{colorcut}) and 47\% and 43\% using the double-Gaussian model 
(equations~\ref{reds}-\ref{fblue}), and they are $\approx 6\%$ lower for the model colors.

We now present our color mark correlation functions.
In practice, in order to obviate redshift-space calculations in the 
halo model and redshift distortions in the data, we use the 
projected two-point correlation function
\begin{equation}
  w_p(r_p)\,=\, \int {\mathrm d}r\,\xi({r_p},\pi)\, 
            = \,2\, \int_{r_p}^\infty \,{\mathrm d}r\,
                         \frac{r\,\xi(r)}{\sqrt{r^2-{r_p}^2}},
\end{equation}
where $r=\sqrt{{r_p}^2+\pi^2}$, $r_p$ and $\pi$ are the galaxy 
separations perpendicular and parallel to the line of sight, and 
we integrate up to line-of-sight separations of $\pi =40\,\mathrm{Mpc}/h$.
We estimate $\xi({r_p},\pi)$ using the Landy \& Szalay (1993) estimator
\begin{equation}
  \xi({r_p},\pi) \,=\, \frac{DD-2DR+RR}{RR},
\end{equation}
where $DD$, $DR$, and $RR$ are the normalized counts of data-data, 
data-random, and random-random pairs at each separation bin.
We then define the marked projected correlation function
\begin{equation}
  M_p(r_p)\,=\, \frac{1\,+\,W_p(r_p)/r_p}{1\,+\,w_p(r_p)/r_p} \, ,
\end{equation}
which makes $M_p(r_p)\approx M(r)$ on scales larger than a few Mpc.
For the SDSS measurements, we used random catalogs with 10 times as 
many points as in the data; the error bars show the variance of 
the measurements of 30 jack-knife sub-catalogs.  

\begin{figure}
 \centering
 \includegraphics[width=\hsize]{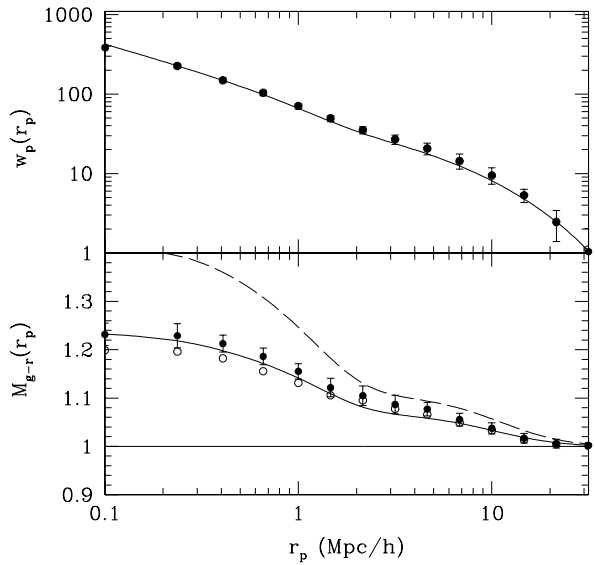} 
 \caption{Projected two-point correlation function and $g-r$ color mark
    correlation function for $M_r<-19.5$.  Points show SDSS measurements 
    for Petrosian (solid points) and model colors (open points), with 
    jack-knife errors.
    Solid curves show the halo-model prediction when satellite galaxies 
    can be drawn from either the red or the blue sequences 
    (equations~\ref{Csatseq}--\ref{pbluesatL}); dashed curve shows the 
    prediction if satellites are drawn from the red sequence only 
    (equation~\ref{reds}).}
 \label{MCF195}
\end{figure}

\begin{figure}
 \centering
 \includegraphics[width=\hsize]{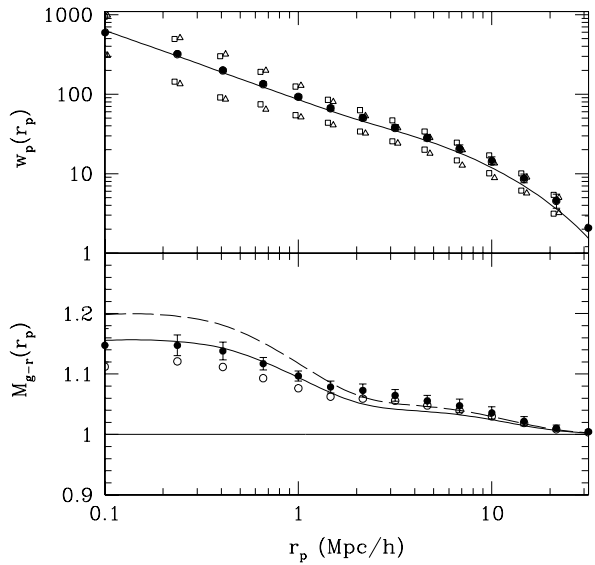} 
 \caption{Projected two-point correlation function and $g-r$ color mark
    correlation function, like Figure~\ref{MCF195}, but for $M_r<-20.5$.
    In the upper panel, the correlation functions for galaxies redder
    and bluer than the color cut (equation~\ref{colorcut}) are also shown,
    for the SDSS galaxies (open squares) and mock catalog galaxies (open 
    triangles).  For clarity, error bars are only shown for the 
    full SDSS catalog.}
 \label{MCF205}
\end{figure}

Figures~\ref{MCF195} and \ref{MCF205} compare the color marked 
correlation functions for the $M_r<-19.5$ and $M_r<-20.5$ catalogs 
with our predictions.  The solid and open points show the 
measurements for Petrosian and model colors.
The color mark signals in the bottom panels are stronger for Petrosian 
colors, at the $1\sigma$ level, for both luminosity thresholds 
$M_r<-19.5$ and $M_r<-20.5$.  Evidently, the environmental 
dependence of Petrosian colors is stronger than that of model 
colors.  However, this is probably due to the fact that the red and 
blue peaks are slightly more displaced from one another for Petrosian 
rather than model colors.

The correlation function of galaxies split by color is the measurement 
that has traditionally been used to show the environmental dependence 
of color (\textit{e.g.}, Zehavi et al. 2005; Tinker et al. 2007).  
The top panel in Figure~\ref{MCF205} shows such measurements for 
galaxies redder and bluer than the color cut given by 
equation~(\ref{colorcut}).  Open squares and triangles are for 
measurements in the SDSS and in a mock catalog constructed as 
described in the previous section.  
(The SDSS galaxies were split by their Petrosian colors; the 
measurement is virtually the same when they are split by model colors.)

The mock catalog is at $z=0$, whereas the SDSS measurements (and 
corresponding theory curves) are at $z\sim0.1$.  Therefore, to compare the 
clustering of red and blue galaxies in our mock with the measurements, 
we measure the ratio of $w_{p,\mathrm{red}}$ to $w_{p,\mathrm{all}}$, and 
$w_{p,\mathrm{blue}}$ to $w_{p,\mathrm{all}}$ in our $z=0$ mock.  We then 
assume that this ratio would be the same at $z=0.1$ as it is at $z=0$; the 
triangles show the result of applying this ratio to $w_{p,\mathrm{all}}$ 
at $z=0.1$ (\textit{i.e.}, the filled circles) --- they represent how the 
clustering of red and blue galaxies differ from the full sample in our mock.
%
The agreement between the clustering of the mock galaxies and SDSS galaxies 
is very good, indicating that our model reproduces these traditional 
measurements of color dependent clustering very well.

Because mark statistics do not require binning of the dataset into 
coarse bins in color, or coarse bins in density, the mark correlation 
functions shown in the lower panels of the figures contain significantly 
more information about environmental correlations than more traditional 
measures.  They allow the mark to take a continuous range in values,
and they yield a clear, quantitative estimate of the correlation
between the mark and the environment at a given scale.
Mark statistics are also sensitive to the distribution of the marks:
for example, for the fainter luminosity threshold ($M_r<-19.5$), 
the color marks have a wider distribution than for the brighter 
threshold, so some galaxies have colors farther from the mean mark.  
Because these outliers also tend to be in more extreme environments, 
the result is a stronger mark correlation 
(\textit{e.g.}, $M_{g-r}(r_p=100h^{-1}\,\mathrm{kpc})\approx 1.23$ 
 \textit{vs.} 1.15).  
Notice also that the mark correlation functions are more curved for the 
fainter luminosity threshold, with a more distinct transition between 
the 1-halo and 2-halo terms.  This is because there are more satellite 
galaxies in the fainter sample, and the mark clustering is more 
sensitive to their spatial distribution within halos.

Note that the model in which no satellites come from the blue sequence 
(dashed curve) produces too strong a signal:  galaxies in dense 
environments are too red.  Since most of these are satellites, this 
model places too many red satellites in massive halos.
The difference between the two models is greater for the faint luminosity 
threshold simply because there are more faint satellites, more of whose 
colors should be drawn from the blue sequence.
The model in which satellites come from a mix of the two sequences, 
though they are increasingly red at large luminosities 
(equation~\ref{Csatseq}--\ref{pbluesatL}) is in good agreement with
the measurements on all scales where the statistic is reliably measured.
This suggests that this model of the colors of central and satellite galaxies is a reasonable one.
The good agreement between our model and the data also indicates
that the correlation between halo mass and environment is the primary 
driver of the environmental dependence of galaxy color.

\section{Discussion}\label{discuss}
We have developed and tested a simple model for several observed 
correlations between color and environment on scales of 
$100h^{-1}\,\mathrm{kpc} < r_p < 30h^{-1}\,\mathrm{Mpc}$.  
Our model is built upon the model of luminosity mark clustering of 
Skibba et al. (2006), in which the luminosity-dependent halo occupation 
distribution was constrained by the observed luminosity-dependent 
correlation functions and galaxy number densities in the SDSS.
The model presented here has added constraints from the bimodal 
distribution of the colors of SDSS galaxies as a function of luminosity.
We make two assumptions: 
 (i) that the bimodality of the color distribution at fixed luminosity 
     is independent of halo mass, and 
(ii) that satellite galaxies tend to follow a particular sequence in 
     the color-magnitude diagram, one that approaches the red sequence 
     with increasing luminosity (equation~\ref{Csatseq}).  
     Alternatively, this assumption can be phrased as specifying how 
     the fraction of satellites which are drawn from the red and blue 
     sequences depends on luminosity (equation~\ref{pbluesatL}).   

One virtue of our model is the ease with which it allows one to 
include color information into mock catalogs.  Adding colors to a 
code which successfully reproduces luminosity dependent clustering 
requires just four simple lines of code --- two for centrals 
and two for satellites (Section~\ref{mock}).  
This is far more efficient than `brute-force' approaches which are 
based on fitting HODs to fine bins in $L$ {\em and} color, or others 
which are based on using observed correlations between color and 
local density.  
Since bimodality is also observed at $z=1$, it would be interesting 
to see if our approach is similarly successful at interpreting the 
measurements of Coil et al. (2008) in the DEEP2 sample.  

Realistic colors are necessary for providing realistic training sets 
for galaxy group- and cluster-finding algorithms, and a number of 
groups are currently developing such mock catalogs.  So we think it 
is worth emphasizing that our approach can be applied to {\em any} 
mock catalog which produces the correct luminosity-dependence of 
clustering.  Thus, although we phrased our discussion in terms of 
an HOD-based mock, mocks based on CLFs or SHAMs could also use our 
method for generating colors.  

In particular, cluster-finding algorithms that exploit information 
about brightest cluster galaxies (BCGs), or galaxies' positions from 
the red sequence, or galaxies' redshift-distorted positions, or the 
multiplicity function or total luminosity or stellar mass of groups, 
could all be tested with mock catalogs constructed with the approach 
described in this paper.  
We will be happy to provide our mock catalogs to those interested, 
upon request.

More generally, we feel that the simplicity of our approach makes it an 
attractive way to begin to include the entire SED into the halo model 
description, and hence into mock catalogs.  Specifically, starting from 
our successful model for adding $g-r$ given $L$, the next step might be 
to add, say, $u-r$, given $g-r$ and $L$ -- again assuming that the 
distribution $p(u-r|g-r,L)$ is independent of halo mass.  This is also 
attractive because we have shown that such an approach is easily 
described using the language of the halo model --- Section~\ref{halomark} 
provides a halo-model description of the color-mark correlation 
function.  This facilitates the use of mark statistics in testing 
our hypothesis that the bimodal color distribution is independent of 
halo mass.  

Comparison of our mark correlation measurements with measurements 
in our mock catalogs and with our halo model calculations
(Figures~\ref{MCF195} and~\ref{MCF205}) suggest that if the 
bimodal color distribution is independent of halo mass, 
then at least some of the noncentral/satellite galaxies in a halo 
must be drawn from the blue sequence --- this fraction of blue 
satellites must be larger at low luminosities.
This is one of the key results of our paper.

If satellites lie on the red sequence because their star formation 
has been quenched by processes such as `strangulation' 
(\textit{e.g.}, Weinmann et al. 2006), then our results suggest that 
quenching is still on-going at lower luminosities.  
Such processes are expected to modify the colors and star formation 
rates of satellite galaxies, but not their morphologies; 
we investigate this further in a subsequent paper by measuring 
morphology mark correlations in the SDSS Galaxy Zoo catalog. 
We caution, however, that we, like all previous halo model analyses, 
have ignored the fact that inclination can affect the observed galaxy 
properties -- luminosities and colors in the present context.  
Corrections for inclination-related effects are available in the 
literature (Giovanelli et al. 1995; Tully et al. 1998; Sheth et al. 2003), 
and they are not negligible.  
Recent work on this by Maller et al. (2008), which appeared while 
our work was being refereed, provides relatively straightforward 
corrections which may be reasonably accurate.  
For this reason, our work should be viewed as attempting a 
halo-model description of the observed colors, rather than providing 
a truly physical picture of the intrinsic (face-on?) colors.  
Of course, if the luminosities and colors had been corrected for 
inclination effects, we expect our analysis to also yield results 
which are closer to the true physical picture.  But because we have 
not yet included these corrections, we believe that statements about 
the physics of `strangulation', especially at low luminosities, are 
premature.  

We expect our model to be in good agreement with the findings of 
Zehavi et al. (2005), who analyzed volume-limited SDSS samples 
after dividing galaxies into two bins in color.  They used a 
slightly redder color cut than did we to produce the measurements 
shown in the top panel of Figure~\ref{MCF205}.  
They found that the fraction of central galaxies which lay 
blueward of this cut increased as $L$ decreased; 
that there were no faint blue satellites; and that, 
although there are blue satellites at intermediate and high $L$, 
they were about a factor of five less common than red satellites 
in halos of the same mass.
Our model is in qualitative agreement, with the mean central and 
satellite galaxy colors increasing with both luminosity and halo 
mass.  
Zehavi et al. inferred from their results that the majority of bright 
galaxies are red centrals of massive halos, and that faint red 
galaxies are predominantly satellites in massive halos.
This is consistent with Swanson et al. (2007), who found 
that both luminous and faint red galaxies are more strongly clustered 
than moderately bright red galaxies.
We reach a similar conclusion, although not all faint red galaxies 
are satellites in massive systems:  some are centrals in underdense 
environments.

We also expect our model to be in qualitative agreement with the 
findings of Blanton \& Berlind (2007). 
These authors defined blue galaxies as those lying blueward of 
$g-r = 0.8 - 0.03(M_r+20)$ (our equation~\ref{colorcut}).  
They then found that the color magnitude relation for galaxies 
in luminous groups tended to have $f_{\rm blue}$ decreasing with 
group luminosity, but that the red and blue sequences were 
otherwise approximately independent of group luminosity.  
They phrased their findings as showing that the color magnitude 
relation depends on group luminosity, presumably because they wished 
to draw attention to the dependence of $f_{\rm blue}$ on group 
luminosity.  In light of the discussion above, we think this is 
slightly misleading.  The red and blue sequences in our model are 
{\em independent} of group properties by construction.  
In our model, the decrease of the blue fraction in luminous groups 
is simply a consequence of the assumption that satellites tend to 
be drawn from the red rather than the blue sequence.  
This happens because more luminous groups will tend to have more 
satellites {\em and} redder centrals (because central galaxy 
luminosity increases with halo mass which is, in turn, strongly 
correlated with total luminosity, and luminous galaxies are red).  
Since our model has mainly red satellites, the red fraction is larger 
in more luminous groups.  Skibba (2008) describes the results of a 
direct comparison of our model predictions with the colors of 
centrals and satellites in group catalogs.  


In our model, \textit{all} environmental correlations arise from the 
fact that massive halos tend to reside in denser environments 
(Mo \& White 1996; Sheth \& Tormen 2002).
Recent studies of the environmental dependence of halo assembly have 
shown that halo properties such as formation time and concentration 
are correlated with the environment at fixed halo mass
(Sheth \& Tormen 2004; Gao, Springel, White 2005; Wechsler et al. 2006;
Croton, Gao, White 2007; Wetzel et al. 2007, Keselman \& Nusser 2007, 
Zu et al. 2007).
They have found that at fixed mass, halos in dense environments form 
at slightly earlier times than halos in less dense environments.
The success of our model suggests that such `assembly bias' effects 
are not the primary drivers of the environmental dependence of galaxy 
colors in the real universe, thus extending previous conclusions about 
the insignificance of assembly bias on galaxy luminosities 
(Skibba et al. 2006; Abbas \& Sheth 2006, 2007; Blanton \& Berlind 2007; 
Tinker et al. 2007), at least for the relatively bright galaxies in 
the SDSS.
Further tests, such as analyses of luminosity and color mark statistics 
of catalogs constructed from semi-analytic models with known assembly 
bias, would shed more light on these issues, and are the subject of a 
subsequent paper.

Our model does not include the galactic `conformity' reported by 
Weinmann et al. (2006), in which bluer centrals are likely to 
be surrounded by bluer satellites, at fixed halo mass.
Including this effect is the subject of work in progress.
The main quantitative predictions of our model, such as the 
\textit{mean} central and satellite colors as a function of mass, 
and the correlations between color and environment, are not expected 
to be significantly affected by this phenomenon, however.
Our model also does not include color gradients within halos --- 
it has long been known that satellite galaxies near halo centers 
tend to be redder than in the outskirts. 
In this case, satellite color marks depend on both the host halo mass 
and on their distance from the halo center.  
Halo model analyses show that this should only matter on small 
scales (see discussion of Fig.~4 in Sheth et al. 2001; Scranton 2002); 
for galaxy populations with many satellite galaxies, the 1-halo term 
of the color mark signal is expected to be slightly higher (Sheth 2005).  
Skibba (2008) incorporates this effect, and does find such an increase 
at small scales.

Finally, it is worth emphasizing that mark statistics are sensitive 
indicators of the correlations between galaxy properties and the 
environment, and as such are powerful tools for constraining galaxy 
formation models.
An analysis of marked correlation with star formation rate marks
in the SDSS and the Millennium Simulation is the subject of work in progress.
The halo-model description of marked statistics, based on the
luminosity dependence of galaxy clustering, also has many applications.
In a forthcoming paper (Skibba \& Sheth 2008), we present a model of 
stellar mass mark correlations and analyze them with SDSS measurements 
analogous to the color mark correlations presented here.


\section*{Acknowledgements}
We thank Xi Kang and Frank van den Bosch for valuable discussions, 
and the Aspen Center for Physics for hospitality in the Summer of 2007, 
where some of this work was completed.
We also thank Tim McKay, the referee, for comments that helped 
to improve the quality of the paper.
This work was supported by NASA-ATP NAG-13720 
and the NSF under grant AST-0520647,
and by HST-AR-10646.

We thank Jeffrey Gardner, Andrew Connolly, and Cameron McBride
for assistance with their {\tt Ntropy} code, which was used 
to measure all of the correlation functions presented here.
$N$tropy was funded by the NASA Advanced Information Systems 
Research Program grant NNG05GA60G.

Funding for the SDSS and SDSS-II has been provided by the 
Alfred P. Sloan Foundation, the Participating Institutions, 
the National Science Foundation, the U.S. Department of Energy, 
the National Aeronautics and Space Administration, 
the Japanese Monbukagakusho, the Max Planck Society, 
and the Higher Education Funding Council for England. 
The SDSS Web Site is http://www.sdss.org/.

The SDSS is managed by the Astrophysical Research Consortium for 
the Participating Institutions. The Participating Institutions are 
the American Museum of Natural History, Astrophysical Institute 
Potsdam, University of Basel, Cambridge University, 
Case Western Reserve University, University of Chicago, 
Drexel University, Fermilab, the Institute for Advanced Study, 
the Japan Participation Group, Johns Hopkins University, 
the Joint Institute for Nuclear Astrophysics, 
the Kavli Institute for Particle Astrophysics and Cosmology, 
the Korean Scientist Group, the Chinese Academy of Sciences (LAMOST), 
Los Alamos National Laboratory, 
the Max-Planck-Institute for Astronomy (MPA), 
the Max-Planck-Institute for Astrophysics (MPIA), 
New Mexico State University, Ohio State University, 
University of Pittsburgh, University of Portsmouth, 
Princeton University, the United States Naval Observatory, 
and the University of Washington.

\appendix

\renewcommand{\thefigure}{\Alph{appfig}\arabic{figure}}
\setcounter{appfig}{1}

\section{Explicit examples of different HODs}\label{zzHOD}
The main text outlines our model; actual implementation of it 
depends on the form of the luminosity-based HOD.  These are of 
two types -- either the relation between halo mass and central 
galaxy luminosity is monotonic and deterministic, or there is 
some scatter.  We use the parametrization of Zehavi et al. (2005) 
to illustrate the former case, and that of Zheng et al. (2007) to 
illustrate the latter.  
The results described in the main text are not particularly 
sensitive to this choice, although the plots we are based on 
HODs in which there is scatter.  


\subsection{No scatter between $L_{\rm cen}$ and halo mass}

To evaluate $n_{\rm sat}/n_{\rm cen}$, suppose that the relation 
between halo mass and central luminosity is deterministic 
(\textit{i.e.}, there is no scatter around the $L_{\rm cen}-m$ 
relation).  Then 
\begin{equation}
 n_{\rm cen}(L) = (dm_L/dL)\, (dn/dm)_{m_L} 
 \label{ncenL}
\end{equation}
where halo model interpretations of SDSS galaxy clustering 
suggest that 
\begin{equation}
 {m_L\over 10^{12}h^{-1}M_\odot} \approx 
 \exp\left({L/1.12\over 10^{10}h^{-2}L_\odot}\right) - 1
 \label{Lcen}
\end{equation}
(Zehavi et al. 2005; Skibba et al. 2006),
and the halo mass function $dn/dm$ is described in Sheth \& Tormen (1999).  

The number density of satellite galaxies which have luminosity $L$, 
whatever the mass of the parent halo, is given by differentiating 
\begin{equation}
 n_{\rm sat}(>L) = \int_{m_L}^\infty dm\, {dn\over dm}\, N_{\rm sat}(>L|m)
 \label{nsatden}
\end{equation}
with respect to $L$.  Zehavi et al. (2005) show that 
\begin{equation}
 N_{\rm sat}(>L|m) \approx \left({m\over 23\,m_L}\right)^{\alpha_L} 
 \label{Nsat}
\end{equation}
where $m_L$ is given by the expression above, and 
\begin{eqnarray}
 \alpha_L &\approx& 1.16 - 0.1(M_r+20) + 0.1 {\rm e}^{-1.5(M_r+21.5)^2}
\end{eqnarray}
is a weakly increasing function of $L$.  
Thus, 
\begin{eqnarray}
 n_{\rm sat}(>L) &=& \left[\exp\left({L/1.12\over 10^{10}h^{-2}L_\odot}\right)
                                       - 1\right]^{-\alpha_L} \nonumber\\
    && \ \times\ 
          \int_{m_L}^\infty dm\, {dn\over dm}\,
              \left({m/23\over 10^{12}h^{-1}M_{\odot}}\right)^{\alpha_L},
\end{eqnarray}
so 
\begin{eqnarray}
 {n_{\rm sat}(L)\over n_{\rm cen}(L)} &=& 
  {1\over 23^{\alpha_L}} + \alpha_L\,{n_{\rm sat}(>L)\over (dn/d\ln m)_{m_L}}
  \nonumber\\
  && \ - \alpha_L\,{d\ln\alpha_L\over d\ln L} \,
            {d\ln L/d\ln m_L\over (dn/d\ln m)_{m_L}}\nonumber\\
  && \ \times \int_{m_L}^\infty dm\, {dn\over dm}\,
    \left({m/23\over m_L}\right)^{\alpha_L}\, 
    \ln\left({m/23\over m_L}\right),
\end{eqnarray}
and the fraction of objects which are centrals is 
\begin{equation}
 f_{\rm cen}(L) = {n_{\rm cen}(L)\over n_{\rm cen}(L) + n_{\rm sat}(L)}
                = {1\over 1 + n_{\rm sat}(L)/n_{\rm cen}(L)}.
 \label{fcen}
\end{equation}
To see what these expressions imply, suppose that $\alpha_L$ 
were independent of $L$.  Then
\begin{equation}
 {n_{\rm sat}(L)\over n_{\rm cen}(L)} = \alpha\, 
  {n_{\rm sat}(>L)\over (dn/d\ln m)_{m_L}} + {1\over 23^\alpha},
\end{equation}
and 
\begin{equation}
 \langle c|m\rangle_{\rm sat} = \langle c|L_{\rm min}\rangle_{\rm red} 
 + \int_{L_{\rm min}}^\infty \!\!\! dL\,C(L)\,
            \left({m_{L_{\rm min}}\over m_L}\right)^\alpha.
\end{equation}
In this case, the mean satellite color is independent of halo mass.  
If $\alpha=1$ (not far off from its actual value) and 
$m\,(dn/dm) \propto \exp(-m/m_*)/m_*$ for some fiducial value 
of $m_*$ (halos more massive than $m_*\approx 10^{13}h^{-1}M_\odot$ 
are indeed exponentially rare), then 
$n_{\rm sat}(>L) = \exp(-m_L/m_*)/(23m_L)$ making 
$n_{\rm sat}(L)/n_{\rm cen}(L) = (m_*/m_L + 1)/23$.  
This ratio decreases as $m_L$ increases---as $L$ increases, 
the ratio of satellites to centrals decreases, and the fraction 
of centrals increases.  

In the analyses which follow, we use the actual halo model values 
of these quantities rather than these approximations.  
A reasonable fit to the actual halo 
model values is given by 
\begin{equation}
  {n_{\rm sat}(L)\over n_{\rm cen}(L)}\approx 
   0.35\, \left[2-{\rm erfc}\Bigl[0.6(M_r+20.5)\Bigr]\right]
\end{equation}
This ratio tends to 0.7 at small luminosities, making the fraction 
of galaxies which are centrals at $L\ll 10^{10}h^{-2}L_\odot$ about 
3/5 (\textit{cf.} equation~\ref{fcen}),
consistent with the satellite fraction $f_\mathrm{sat}(L)$ of 
van den Bosch et al. (2007a).

\subsection{Stochasticity in the $L_{\rm cen}-m$ relation}

Zheng et al. (2007) allow for stochasticity in the relation between 
halo mass and central galaxy luminosity.  They assume that 
\begin{equation}
  P(\mathrm{log}\,L_\mathrm{cen}|M) 
  = \frac{1}{\sqrt{2\pi}\sigma_{\mathrm{log}L}}
    \mathrm{exp}\Biggl[-\frac{[\mathrm{log}(L_\mathrm{cen}/\langle L_\mathrm{cen}|M\rangle)]^2}{2\sigma_{\mathrm{log}L}^2}\Biggr] ,
\end{equation}
and then set 
\begin{equation}
  \langle N_\mathrm{cen}|M\rangle \,=\, \frac{1}{2}\Biggl[1\,+\,\mathrm{erf}\Biggl(\frac{\mathrm{log}(M/M_\mathrm{min})}{\sigma_{\mathrm{log}M}}\Biggr)\Biggr]
\label{NcenM}
\end{equation}
and
\begin{equation}
  \langle N_\mathrm{sat}|M\rangle = 
    \Biggl(\frac{M-M_0}{M_1^{ ' }}\Biggr)^\alpha .
\label{NsatM}
\end{equation}
The Poisson model for satellite counts sets 
\begin{equation}
  \langle N_\mathrm{sat}(N_\mathrm{sat}-1)|M\rangle 
  = \langle N_\mathrm{sat}|M\rangle^2.
\label{NsatMsq}
\end{equation}
Their Table~1 shows how all of the parameters in this HOD vary with 
SDSS $r$-band luminosity.  We have found that these scalings with 
$L_r$ are well approximated by 
\begin{eqnarray}
\frac{M_\mathrm{min}}{10^{11.95}M_\odot/h} &\approx&
  \mathrm{exp}\Biggl(\frac{L}{10^{10.0}L_\odot/h^2}\Biggr)-1 \\
\sigma_{\mathrm{log}M} &\approx&
  \left\{ \begin{array}{ll}
    0.26 \,\mathrm{if}\, M_r>-20.5 \\
    0.385-0.25\,(M_r+21), \,\mathrm{otherwise} 
  \end{array} \right. \\
M_1^{ ' } &\approx& 17\,M_\mathrm{min} \\
\frac{M_0}{10^{11.75}M_\odot/h} &\approx& 
  \Biggl(\frac{L}{10^{9.9}L_\odot/h^2}\Biggr)^{0.6} \\
\alpha &\approx& 1 \,-\, 0.07\,(M_r+18.8)
\end{eqnarray}
As in Zehavi et al. (2005), the value of $M_1^{ ' }/M_\mathrm{min}$, which 
determines the critical mass above which halos typically host at least one 
satellite galaxy, is approximately independent of luminosity, while the 
$\langle N_\mathrm{sat}\rangle$ slope $\alpha$, which characterizes the mass 
dependence of the efficiency of galaxy formation, increases with luminosity.
The two new HOD parameters are $\sigma_{\mathrm{log}M}$ and $M_0$.  
They are not constrained well and their uncertainties are large 
(see Zheng et al. for details), but our correlation functions and 
color mark correlation functions are not very sensitive to their exact 
values.

For the two luminosity thresholds discussed in the main text, 
$M_r<-19.5$ and $M_r<-20.5$, the parameters above are  
$M_\mathrm{min}=5.8\times10^{11}\,h^{-1}\mathrm{Mpc}$ and 
$M_\mathrm{min}=2.2\times10^{12}\,h^{-1}\mathrm{Mpc}$, 
and the effective value of $M_1/M_\mathrm{min}\approx20$, 
approximately independent of luminosity, 
is similar to the factor of 23 in the Zehavi et al. HOD.

For our purposes, the main difference with this HOD model is the 
scatter in luminosity at fixed mass.
We first discuss how to construct a mock catalog that includes this 
scatter.  We then explain how our model of the color mark is modified. 

To account for the scatter between $L_\mathrm{cen}$ and 
$M_\mathrm{halo}$ in the mock catalogs, we do not simply select 
the subset of halos in the simulation which have 
$M>M_\mathrm{min}(L_\mathrm{min})$, as we do in the case of a 
sharp threshold (e.g. Section~\ref{mock}).
Instead, we generate uniformly distributed random numbers $u$ 
between 0 and 1 for each halo of mass $M$.  Then we keep the halo if 
$u\,<\,\langle N_\mathrm{cen}|M\rangle$ (equation~\ref{NcenM}).  
As a result, only half of the halos with $M\approx M_\mathrm{min}$ 
are kept, as are quite a few halos with $M<M_\mathrm{min}$.  
Larger values of $\sigma_{\mathrm{log}M}$ increase the range of 
halo masses around $M_\mathrm{min}$ and increase the total
number of halos because the abundance of halos increases with 
decreasing mass.

Our halo model of the color mark is also modified by the scatter between
luminosity and mass, and hence $\langle N_\mathrm{cen}|M,L_\mathrm{min}\rangle$
is no longer a step function.  The central galaxy color mark, described in
Section~\ref{model}, is slightly more complicated.  
The mean central galaxy color as a function of luminosity 
$\langle c|L\rangle_\mathrm{cen}$ (equation~\ref{Ccen}) depends on 
the number density of central galaxies as a function of luminosity
$n_\mathrm{cen}(L)$ (equation~\ref{ncenL}), which now includes an 
integral:
\begin{eqnarray}
  n_\mathrm{cen}(L) &=& \frac{d}{dL}\,n_\mathrm{cen}(>L) \nonumber \\
    &=& \biggl(\frac{dn}{dM}\biggr)_{M_L} \biggl(\frac{dM_L}{dL}\biggr) \\
    &\phantom{=}& +\, \int dM\,\frac{dn}{dM}\,\frac{d}{dL}
      \langle N_\mathrm{cen}|M,L_\mathrm{min}\rangle \nonumber
\end{eqnarray}
Then the central galaxy color mark, which is used in the color 
mark correlation functions, is also an integral (equation~\ref{CcenM}):
\begin{equation}
  \langle c|M\rangle_\mathrm{cen} \,=\,
    \int dL\, P_\mathrm{cen}(L|M)\,\langle c|L\rangle_\mathrm{cen}.
\end{equation}
The model of the color mark correlation functions, described in 
Appendix~\ref{colorMCF}, is also modified.  
However, we reiterate that, in general, the correlation functions and 
color mark correlation functions are not sensitive to the exact amount 
of scatter in mass at fixed luminosity.


\section{A halo-model of color mark correlations}\label{colorMCF}


We perform our halo model calculations in Fourier space.  
The two-point correlation function is the Fourier transform of the 
power spectrum
\begin{equation}
  \xi(r) = \int {dk\over k}\, {k^3 P(k)\over 2\pi^2}\, {\sin kr\over kr}.
 \label{xiFT}
\end{equation}
In the halo model, $P(k)$ is written as the sum of two terms:
one that arises from galaxies within the same halo and dominates on 
small scales (the 1-halo term), and the other from galaxies in 
different halos which dominates on larger scales (the 2-halo term).  
That is,
\begin{equation}
  P(k) = P_{1h}(k) + P_{2h}(k),
 \label{1h2h}
\end{equation}
where,
\begin{eqnarray}
  P_{1h}(k) &=& \int dM \,\frac{dn(M)}{dM}\,
                \langle N_{\mathrm {cen}}|M\rangle \nonumber \\
	&& \ \times \Biggl[ \frac{2\,\langle N_{\mathrm {sat}}|M\rangle\,
	   u_{\mathrm {gal}}(k|M)}{\bar n_{\mathrm {gal}}^2} \nonumber \\
       &\phantom{=}& \qquad +\, 
           \frac{{\langle N_{\mathrm {sat}}(N_{\mathrm {sat}}-1)|M\rangle}\,
	  u_{\mathrm {gal}}(k|M)^2}{\bar n_{\mathrm {gal}}^2}  \Biggr], 
  \label{Pk1h} \\
  P_{2h}(k) &=& \Biggl[ \int dM \,\frac{dn(M)}{dM}\,
                 \langle N_{\mathrm {cen}}|M\rangle
  \label{Pk2h} \\
       &\phantom{=}& \;\; \times \, \frac{1\,+\,
          \langle N_{\mathrm {sat}}|M\rangle\,
	  u_{\mathrm {gal}}(k|M)}{\bar n_{\mathrm {gal}}}\,b(M) \Biggr]^2 \,
	  P_{\mathrm {lin}}(k), \nonumber
\end{eqnarray}
where the number density of galaxies $\bar n_{\mathrm {gal}}$ is 
(\textit{cf}., eq.~\ref{nsatden})
\begin{equation}
  \bar n_{\mathrm {gal}} = \int dm \, {dn(m)\over dm}\, 
    \langle N_{\mathrm {cen}}|m\rangle\,
    \Bigl[1 + \langle N_{\mathrm {sat}}|m\rangle\Bigr]
 \label{ngal}
\end{equation}
and $u_{\mathrm {gal}}(k|M)$ is the Fourier transform of the galaxy 
density profile.  It is standard to assume this has the same form as 
the dark matter, so we use the form for $u$ given by Scoccimarro et al.(2001).
The distribution $p_{\rm sat}(N_{\rm sat})$ is expected to be 
well-approximated by a Poisson distribution (\textit{e.g.}, Kravtsov et al. 2004; 
Yang et al. 2008), so we set 
$\langle N_{\mathrm {sat}}(N_{\mathrm {sat}}-1)|M\rangle\,
 =\,{\langle N_{\mathrm {sat}}|M\rangle}^2$.
The two parts of the 1-halo term in equation~(\ref{Pk1h}) can be 
thought of as the `center-satellite term' and the 
`satellite-satellite term'.

To describe the effect of weighting each galaxy, we use $W(k)$ to 
denote the Fourier transform of the weighted correlation function.
Like the power spectrum, we write this as the sum of 1- and 2-halo 
terms:
 $W(k) \,=\, W_{1h}(k) \,+\, W_{2h}(k)$.
Since central and satellite galaxies have different properties, 
we weight central and satellite galaxies separately by their mean 
mass-dependent marks: $\langle c|m\rangle_{\rm cen}$ and 
$\langle c|m\rangle_{\rm sat}$ (Section~\ref{model}).  
Following Sheth (2005), we write 
\begin{eqnarray}
  W_{1h}(k) &=& \int dM \,\frac{dn(M)}{dM}\,
                 \langle N_{\mathrm {cen}}|M\rangle \nonumber\\
    && \, \times\ \Biggl[ \frac{2\,c_{\mathrm {cen}}(M)\,
                       \langle c_{\mathrm {sat}}|M,L_{\mathrm {min}} \rangle\,
	     \langle N_{\mathrm {sat}}|M\rangle\,u_{\mathrm {gal}}(k|M)}
	     {\bar n_{\mathrm {gal}}^2 \,\bar c^2} \nonumber\\
	   && \,+\, \frac{ {\langle N_{\mathrm {sat}}|M\rangle}^2 \,
		{\langle c_{\mathrm {sat}}|M,L_{\mathrm {min}}\rangle}^2\,
                 u_{\mathrm {gal}}^2(k|M)}
		 {\bar n_{\mathrm {gal}}^2 \,\bar c^2} \Biggr],
  \label {Wk1h}\\ 
  \frac{W_{2h}(k)}{P_\mathrm{lin}(k)} 
	       &=& \Biggl[ \int dM \,\frac{dn(M)}{dM}\,
                   \langle N_{\mathrm {cen}}|M\rangle\,b(M) \label{Wk2h} \\ 
		&& \frac{c_{\mathrm {cen}}(M)\,+\,\langle N_{\mathrm {sat}}|M\rangle\,
		  \langle c_{\mathrm {sat}}|M,L_{\mathrm {min}}\rangle\,
                  u_{\mathrm {gal}}(k|M)}
		{\bar n_{\mathrm {gal}}\,\bar c} \Biggr]^2 \, ,
		\nonumber
\end{eqnarray}
where we normalize by the mean color mark
\begin{eqnarray}
  \bar c &=& \int dM \,\frac{dn(M)}{dM}\,
            \langle N_{\mathrm {cen}}|M\rangle \nonumber\\
	&&\qquad\times\quad 
  \frac{c_{\mathrm {cen}}(M)\,+\,\langle N_{\mathrm {sat}}|M\rangle\,
		  \langle c_{\mathrm {sat}}|M,L_{\mathrm {min}}\rangle}
	   {\bar n_{\mathrm {gal}}} \,.
 \label{cbar}
\end{eqnarray}


\label{lastpage}

\end{document}